\documentstyle[12pt]{article}
\begin{document}
\baselineskip=22pt plus 0.2pt minus 0.2pt
\lineskip=22pt plus 0.2pt minus 0.2pt

\begin{flushright}
Los Alamos preprint \\
LA-UR-94-2634\\
hep-ph/9410315
\end{flushright}

\vspace*{0.2in}

\begin{center}
\large

Confinement, Crossing Symmetry, and Glueballs \\

\vspace*{0.5in}

Alfred S.\ Goldhaber$^{1),2),3)}$ and T.\ Goldman$^{1)}$

\end{center}

\vspace*{0.25in}
$^{1)}$ Theoretical Division, Los Alamos National Laboratory, NM 87545,
USA

$^{2)}$ Physics Division, Los Alamos National Laboratory, NM 87545, USA

$^{3)}$ Institute for Theoretical Physics, State University of New York,
Stony Brook,
NY 11794-3840, USA

\vspace*{0.25in}

\begin{center}
\large

\vspace{0.5in}

\today

\vspace{1in}

ABSTRACT

\end{center}

\noindent It is suggested that the quark-confining force is related by
crossing symmetry to a color-singlet glueball ${\cal G}$ which is well
described as a loop of one quantum of color magnetic flux.  Electron
pair annihilation as high as $\approx 2 GeV$ above the $\Upsilon$ mass
could produce $\Upsilon \rightarrow \ell^+\ell^-$ accompanied by ${\cal
G}$ or one of its excited states.
\newpage

\noindent{\bf I. Introduction}

The concept of crossing symmetry has a venerable place in the history
of particle physics.  At least since Yukawa,  forces between particles
have been viewed as derived from the (virtual) exchange of other
particles, which themselves can be produced as real, observable objects
by acceleration of the particles which the forces influence. If the
colliding particles carry appropriate charges, then the force-carrying
objects also may be produced by annihilation, but the acceleration or
`bremsstrahlung' mechanism is the generic exemplar of crossing
symmetry.  Our goal here is to explore implications of adopting that
same viewpoint to describe the confining force between quarks, a force
whose existence is indicated both by phenomenological analyses of
hadronic (and especially heavy quarkonium) spectra ${\cite{fenomconf}}$
and by lattice calculations in pure QCD without dynamical
quarks ${\cite{latconf}}$. Both types of analysis are consistent with
a dominantly Lorentz scalar potential whose magnitude grows linearly
with separation between heavy quark and antiquark ${\cite{recent}}$.

\setcounter{page}{1}

Before seeking a candidate for the exchanged particle, let us try to
classify the properties it would require. First, it need not be coupled
to a charge carried by gluons, since the charge of any color octet
object can be screened by the creation of gluon pairs. Thus, even in
the absence of light quarks, one would not expect at large distances to
find a (linear) confining potential between a heavy color octet
particle and its antiparticle. Secondly, one expects the force to bind
a quark to an antiquark, but also to bind two quarks, as part of the
binding of three quarks to make a baryon.   Thus it is permitted, and
therefore in our view a desirable simplification, to assume that the
exchanged object is a color singlet whose coupling to a particle
depends at most on the color $SU(3)$ representation to which that
particle belongs.  Finally, since the force should be attractive, the
simplest possible and therefore most appealing spin assignment is spin
zero.  Of course this might well be only the lowest in a hierarchy of
spin states, as in the familar example of Regge theory ${\cite{landshoff}}$.

If for the moment we continue to ignore light quarks, then the object
must be constructed out of the only available degrees of freedom, i.e.,
gluons.  Such an object is by definition a glueball.  There is a great
deal of literature on the possible structure of (color singlet)
glueballs ${\cite{gb}}$, and on whether some experimentally observed
meson resonances may be identified as glueballs ${\cite{gbexp}}$. Up
to the present such studies have been inconclusive.  Current estimates
of masses for $0^{++}$ glueballs lie above $1.5 \, GeV$, and for
$2^{++}$, above $2 \, GeV$ ${\cite{gb}}$.

Since confinement and glueballs are two of the most elusive and
difficult phenomena conjectured to occur in the strong interactions, it
is not obvious that trying to consider them together is advantageous.
Our justification for doing so rests on the outcome -- suggestions for
new laboratory and lattice experiments to be described below.

It is implausible to model a glueball as a collection of some definite
number of gluons, since gluon number is not a well-defined quantity at
low energy and large length scales.  Already at order $g_S(s)$ (where
$g_S(s)$ is the strong coupling constant at the appropriate scale $s$),
there are Feynman graphs for an initial single gluon to change to a
final gluon pair, making it hard to see how a resonant state could be
described in terms of a fixed number of gluons.

Given the intrinsic imprecision of gluon number at large length scales,
a glueball might best be pictured in first approximation as a
spread-out, classical gluon field configuration, which therefore would
have a strong form factor suppression of its couplings at large
momentum transfers, corresponding to small distance scales. Clearly
this would imply a suppression of glueball production by
quark-antiquark annihilation, making such objects difficult to observe
cleanly in ordinary hadronic processes. In the absence of light quarks
the lightest glueball should be completely stable, but in practice
there is no easy way to distinguish a glueball from a color and flavor
singlet combination of quark and antiquark.  Thus the lack of
unambiguous experimental evidence for glueballs should not be
surprising.

If the object does not carry color, then what couplings might be
possible?  Since it is supposed to be made of gluons and to confine
quarks, its coupling must be a non-trivial function of color.   We have
speculated above that there may be no confining potential for color
octets, which of course would mean that the coupling of our exchanged
object to octets would vanish.  Assuming this, the only allowed form of
coupling would be a function of the set of operators in the center of
color $SU(3)$, namely the triality, with eigenvalues $T = -1, 0, +1 $.
This is zero for multiplets in the adjoint (octet) representation and
representations which can be constructed from it, and $N (mod\, 3)$ for
representations made of $N$ quarks.  For a specific form of coupling to
an object $X$, we suggest $g_{{\cal G}XX}=g_0sin \, ^{2} \, (\pi \!  <
\!T \! > \! \! /3)$, where $< \!T \! >$ is the expectation value of $T$
for a volume of size equal to the geometrical dimensions of a glueball.
This has the requisite periodicity in $< \!T \! >$ and is always of one
sign, so that it can never lead to repulsive interactions.  For an
isolated heavy quark, if light quark contributions are ignored, then
the expectation value $< \!T \! >$ should be an eigenvalue, as gluon
fields cannot change $T$.

It is readily seen that our proposed glueball exchange gives exactly
the same (scalar) attraction between two quarks as between a quark and
an antiquark. It also would give the same attraction between a quark
and a diquark (whether $\bar3$ or $6$), when the constituents of the
latter were closer together than the characteristic glueball size.
Note that this contrasts with the characteristics of the perturbative,
Coulombic octet exchange force between quarks, which does differ for
different objects with the same triality.

We have arrived at a picture in which the schematics of confinement
forces lead by crossing symmetry to properties of a scalar,
color-singlet glueball which ought to be hard to produce or destroy in
annihilation reactions, and which is weakly coupled to ordinary
color-singlet matter. The internal structure of the glueball suggests a
likely tower of excitations, all with quite small widths, since any
familiar hadron systems light enough to be radiated in transitions from
one ${\cal G}$ state to another should have small coupling to this
exotic object.  For the same reason, the mass of the lightest ${\cal
G}$ state must lie quite far above the threshold for a system of two
$\pi$ mesons, since otherwise there would be a visible sharp state or
states in the dipion spectrum.  This is a complementary viewpoint to
that of Novikov et al. ${\cite{NOV}}$, who, on the basis of QCD sum rules,
suggested studies of the dipion spectrum perhaps due to a scalar
glueball in $\psi \rightarrow 2\pi + \gamma$.

\noindent{\bf II. Perturbative approach}

Let us turn to a perturbative approach to make the discussion more
specific.  In perturbation theory the coupling due to gluon exchange
between two quarks which are off mass shell by some characteristic
amount $\Delta M$ may be estimated at small momentum transfer $q$ by
focusing on the most singular part of the QCD coupling.

For exchange of the two-gluon color singlet combination, the
quark-quark potential in momentum space should be
$$
\tilde{V}(q) \approx \frac{(\alpha_{s}(q))^{2}q^4}{(\Delta M)^2q^4}, \eqno(1)
$$
where $\alpha_S(q)$ is $g_S^2(q)/4\pi$, the factor of $q^4$ in the
numerator comes from the integration over (small) loop momentum, the
$q^4$ in the denominator comes from the two gluon propagators, and
$(\Delta M)^2$ from the quark propagators.  Let us assume the
Richardson ${\cite{rich}}$ ansatz for the leading behavior of
$\alpha_{s}$,
$$
\alpha_s(q) = \frac{12 \pi}{(33 - 2 n_{f}) ln(1 + q^2/\Lambda^{2})}, \eqno(2)
$$
where $\Lambda$ is of order the QCD scale but not necessarily equal to
$\Lambda_{\overline{MS}\,}$, and the coefficient is determined by the
one-loop $\beta$-function for QCD which depends on the number of light
quark flavors, $n_{f}$.  This implies a pole in $\alpha_{s}$ at $q^{2}
= 0$, giving a double pole in $\tilde{V}(q)$ and hence by Fourier
transformation a linearly rising potential $V(r)$ in coordinate space.

It is straightforward to check that if arbitrary numbers of gluons are
exchanged they again lead to a linearly rising potential, as the
additional powers of $q$ from the propagators and vertices cancel
against those from the additional loop integrals.  Thus we may consider
the glueball exchange as a sum over all numbers of gluon exchange,
which is as near as one might hope to approach a nonperturbative result
in a perturbation theory framework.  If $\Delta M$ is of the same order
as $\Lambda$, then all numbers of exchanges should contribute
comparably, and the possibility of obtaining a nonpolynomial coupling
(such as that to triality suggested above) also becomes
comprehensible.

The inclusion of arbitrary numbers of gluon exchanges in principle
allows arbitrary spin exchange, as would occur in Regge theory.  This
raises the possibility of coupling the spins of the quarks.  However,
that would lead to extra factors of $q$ in the numerator, and so would
not contribute to the confining potential.

\pagebreak

\noindent{\bf III. Extrapolations}

If in the lowest-order description the exchanged object responsible for
a color-singlet linearly rising potential is a digluon, we also should
look at the digluon in the crossed channel to get a hint about the
properties expected for the glueball.  We can see a serious difficulty
at once:  We have some insight near $q^2 = 0$, but that is a point well
below threshold in the heavy quark-antiquark channel.  Nevertheless,
let us try. From standard parity considerations, we know the quarks
must be in a relative P-wave to produce a scalar glueball.  Such a
state would have nonzero overlap with configurations corresponding to
opposite-pointing magnetic moments for the quark and antiquark, which
of course are on opposite sides of the center of mass.  This suggests a
field configuration which is a loop of color magnetic flux.

There is much evidence that the confining effect should be viewed as
due not to a mere static scalar potential, but rather to an electric
flux string stretching from quark to antiquark ${\cite{isgurefs}}$.
With confinement described this way, it might seem natural to think
that glueballs should be described as loops of that electric flux.
However, there are reasons to question whether electric loops could
play a distinctive role, with some unique signal of their presence.
First, imagine an electric loop in isolation.  A spontaneous
fluctuation breaking the loop by creation of a light quark pair could
`unzip' the loop at the speed of light.  This is the `hybrid' problem
${\cite{barnes}}$: It is effectively impossible to distinguish an
electric flux loop of any size from a collection of one or more
mesons.

Difficult as they might be to observe, electric flux loops in principle
could be radiated by accelerated quarks. To decide what quanta are
exchanged to produce this force, imagine giving a sudden large impulse
to the quark at one end of the hypothesized electric flux string.  The
most likely result will be to break the string in one or more places,
producing a number of quark-antiquark mesons.  To produce an electric
flux loop, the impulse might put the string into an excited state where
it has a kink.  This can break off leaving an intact string plus a
loop, which again would decay quickly to ordinary mesons, leaving no
characteristic trace.

To convert a large loop of color $\underline {magnetic}$ flux into
quarks would require lining up many quark-antiquark pairs head-to-tail,
each with magnetic moment parallel to the local direction of the loop.
Thus such a gluon field configuration would have small overlap with
typical quark configurations, and should be minimally affected by the
existence of light quarks.  Magnetic flux is proportional to magnetic
moment per unit length, so the number of pairs would be proportional to
the length of the loop.  Since such a quark configuration is quite
improbable, as it involves many correlated pairs, the coupling of large
magnetic flux loops to conventional mesons should be quite weak.

There is a certain appeal in identifying magnetic loops as the true
partners of a force pictured in terms of an electric flux string, since
crossing symmetry interchanges temporal and spatial directions, and
thus ought to interchange electric and magnetic fields.  However, our
main point is that the magnetic type should be distinguishable from
conventional hadrons, unlike the electric type.

Still another reason why one might expect magnetic loops is found in
the Copenhagen ``flux spaghetti'' picture ${\cite{nielsen}}$. There the
elemental fluctuations of the pure glue QCD vacuum include tubes of
magnetic flux, so that one expects the low-lying excitations to include
loops of flux and possibly color magnetic monopoles.  If we think about
the separation of the heavy quarks as analogous to the separation of
ordinary magnetic monopoles inside a superconductor, then by the
analogy we might expect generation in our problem of a loop of color
magnetic current.  Whether there is a distinction in this nonabelian
system between magnetic current and magnetic flux is unclear to us.  In
the following we proceed as if there is no difference.

In summary, crossing symmetry almost certainly holds for the long-range
force between heavy quarks, but the naive crossing-symmetric partner, a
color electric flux loop, would be practically indistinguishable from
ordinary meson systems.  However, if large color magnetic loops were
produced, their coupling to light quarks might be weak enough to allow
distinctive signals of their presence.  We  want now to suggest that
exactly such an effect could occur in a simple and natural way.

\noindent{\bf IV. Experimental and theoretical tests}

Consider an $e^+e^-$ reaction which produces a $b\bar b$ pair at a
center of mass energy near or above the threshold for free $B$ meson
pair production.  Let us attempt to picture the evolution of the system
by naive classical electrodynamics, waiting until later to impose
quantum constraints on the description.  Ignoring light quarks for the
moment, we know that as the $b$ quarks separate by about $1 f \! m$
they will increase in their effective color electric charge coupling to
a value of order unity.  If we assume the relevant 4-current is locally
conserved, then it follows that there will be a large spatial current
pulse across the midplane between the quarks.  This current will induce
a color magnetic flux loop with flux also of order unity. However, once
the current pulse has died down, the flux loop (if it approaches
sufficient mass to be a glueball) should no longer be static and so
should begin to execute motions with one or more periods (perhaps like
a TE mode in a microwave cavity, but with the roles of electric and
magnetic fields interchanged).

Meanwhile, the heavy quarks will have lost energy, perhaps enough to
put them below threshold for $B$-meson pair production.  Eventually
they will be forced by the linear potential to stop separating and then
to come back together.  Again, as they pass through $1 f \! m$
separation, an oppositely directed pulse of current will be generated
across the midplane.  Depending on how the original magnetic loop is
configured by this stage, the new loop induced by the second current
pulse may tend to cancel the first one, or double its field strength
and quadruple its stored energy, or something in between.  If the
resulting gluon field configuration is metastable (i.e, a long-lived
resonant state ${\cal G}$), and the leftover quark energy is close to
the $\Upsilon$ mass, there should be a high probability of forming the
$\Upsilon$ state.  This state in turn could be identified reliably by
its decay to a lepton pair.

The key idea here is that the slowly moving heavy quarks could generate
a coherent, approximately classical gluon field configuration.  Because
of its weak coupling to ordinary hadron channels, this configuration
might not decay before the heavy quarks have settled into a low
`bottomonium' state, which has a finite branching fraction for decay
into leptons. In those leptonic decay events, contamination of the
glueball structure by ordinary mesons would be minimized.  The process
is a close analogue to photon emission in a positronium atomic
transition, and thus justifies the invocation of  crossing symmetry to
describe it.  As far as we are aware, this reaction is unique among
those accessible to laboratory observation in offering the possibility
of ${\cal O} (1)$ coupling between quarks and `solitons' of the gluon
field. The reason is that the heavy quarks move fairly slowly, so that
creation of light quark pairs might be suppressed. Note that the loop
of flux must contain one quantum (meaning that a quark which circled
the flux would suffer an Aharonov-Bohm phase shift $\pm 2\pi/3$) if it
is to be a color singlet, able to decouple from the heavy quarks whose
motions produced it.

Having suggested an experimental test of our general scheme, we also
wish to propose a theoretical test of the particular claim that the
linear confining potential may not affect gluons.  Consider a net color
singlet made up of two heavy octets and one heavy {\bf 27}
representation of color $SU(3)$, with their locations specifying the
vertices of a triangle.  The short-range Coulomb interaction between
the two octets is repulsive, since together they make a {\bf 27}, but
if they are separated sufficiently their interaction should be screened
by gluons, with no dependence of the static energy on further
separation.  If there were a linear confining force, one might expect
the potential to show a minimum at intermediate separation, while
without such a force the potential should be monotonically decreasing.
By studying the energy of the whole color-singlet system while varying
the length scale as well as the two angles which define the triangle,
one may be able to check whether an explicit confinement force is
necessary for simple fits to the energy dependence.

\noindent{\bf V. Confronting the hybrid problem}

If the total energy of the $b\bar b$ pair were well above $B$-meson
production threshold, then the conventional expectation would be that
conjugate light quarks accompany each of the heavy quarks, suppressing
the long range color separation which we propose as the generator of a
color magnetic flux loop.  Even below threshold, light quark pair
production in principle competes with color magnetic loop production.
If we accept typical estimates of minimum glueball masses in the 1 GeV
region, then ${\cal G}$ production could not occur except above $B\bar
B$ threshold, where $B$-meson production competition must be strong.
Nevertheless, study of mesonic states accompanying final states with
$\Upsilon$ decay to a lepton pair could well give a much cleaner view
of the glueball channel than conventional hadron collisions.  Our
suggestion is that there should be a nonzero probability of generating
the magnetic loop instead of the light quark pair, thus taking away so
much energy from the heavy quarks that they collapse to the $\Upsilon$
ground state.

Let us estimate the hadronic background to this signal.  The 3S state
has an inclusive branching ratio $\leq 0.1$ for producing a 1S
$\Upsilon$.  Therefore the 4S state, with a width $10^3$ times greater,
should have a branching ratio to $\Upsilon$ of $\leq 10^{-4}$, implying
a background dilepton branching ratio signature $\leq 10^{-6}$.

Another background is the purely electromagnetic process $e^+e^-
\rightarrow \mu^+\mu^-2\gamma$, which intrinsically is down by order
$\alpha^2$ compared to $B\bar B$ production, with the final
$\mu^+\mu^-$ mass distributed smoothly over a large range, so that the
background under the $\Upsilon$ signal should be utterly negligible.  A
process with one final $\gamma$ can be excluded by a cut on the center
of mass momentum of the final lepton pair.

The crucial issue now is the size of the ${\cal G}$ signal, i.e., the
expected rate for production of a $\mu^+$ and $\mu^-$ with pair mass
equal to that of an $\Upsilon$.  Here our reasoning is quite simple.
We argue for an ${\cal O} (1)$ effect once the glueball production
threshold is surpassed.  Consider, for example, a $1/N_c$ argument
(where $N_c$ is the number of colors in the strong gauge theory):  In
the large $N_c$ limit,  the glueball production branching ratio should
be suppressed by a factor ${\cal O} (1/N_{c}^2)$, compared to the
\underline {inclusive} rate for $b \bar b \rightarrow all$.  This
implies an order of magnitude suppression of the proposed reaction
compared to the overall rate for background processes not involving a
glueball (in this setting mainly $B$-meson pair-production).  However,
these dominant processes would not particularly populate the region of
phase space in which the two heavy quarks are found in the $\Upsilon$
ground state.  Consequently, conventional hadronic backgrounds should
not be expected to obscure a glueball signal as they are also suppressed
in this (or any given) region.  Thus, while it is easy to envision
factors which could reduce the effect by an order of magnitude, if
it were down by more than two orders of magnitude (in rate),
identification as a strong-coupling phenomenon would not be credible.

Therefore, since the decay $\Upsilon \rightarrow \mu^+\mu^-$ has a
branching ratio $\geq 0.01$ for each of the first three ($S$-)states,
we expect that the branching ratio (in comparison to the $B\bar B$
production rate) for the $\mu^+\mu^-$ signal after $e^+e^-$ collision
at the appropriate energy should lie between $10^{-4}$ and $10^{-2}$.
This should be compared with the conventional $\Upsilon$ cascade
background estimated above, which is at least $100$ times smaller.  If
a rise in the observed ratio is seen as the initial $e^+e^-$ energy
increases beyond some value above $B\bar B$ threshold, then it becomes
interesting to look for peculiarities in the accompanying hadron
configurations.  As the initial collision energy rises further, it
might become possible for ${\cal G}$ emission to populate excited
$\Upsilon$ states, which could be observed directly by their own
leptonic decay, at the possible risk of contamination by cascade
background more serious than for the $\Upsilon$ ground state.

What are the most promising choices of $e^+e^-$ collision energy for
this experiment?  If we consider the process as analogous to a
radiative atomic transition, then the obvious choice is to produce one
of the known resonant states of the $b\bar b$ system, so that the
${\cal G}$ emission channel, once open, can compete with $B\bar B$
production.  The radiation should have a form factor with
characteristic ${\cal G}$ momentum width $\Delta p \approx 200 \,
MeV/c$.  Given the uncertainty in ${\cal G}$ mass, $\Upsilon$ states
with excitation energies up to and even beyond $2 \, GeV$ should not be
excluded from the search.

Finally, having referred to possible narrow excited states of ${\cal
G}$, we should discuss their expected spacing.  Assuming a mass of
order $1 \, GeV$ and radius of order $1 f \! m$, one obtains for the
inverse moment of inertia of a loop about an axis through its plane the
value ${\cal I}^{-1} = 80\, MeV \! \! ,$  which by the rule
$E=J(J+1)/2{\cal I}$ gives $80 \, MeV$ for the energy of the first
excited state.  (If the radius were even larger and so the spacing
smaller, detector resolution limitations could lead to the appearance
of a broader state, misshapen from a standard Lorentzian.) It would be
interesting if narrow states with this rotational-band splitting could
be distinguished experimentally, as they have not been detected in
lattice calculations ${\cite{gb}}$.

\noindent{\bf VI. Conclusions}

We have proposed an analogue for the confinement force of Yukawa's
$\pi$ meson for the nuclear force.  Our candidate is a glueball state
${\cal G}$, a loop of color magnetic flux whose existence might be
verified and more detailed properties studied by electron-positron
inelastic collisions leading to $\Upsilon \rightarrow \mu^+\mu^-$ plus
(possibly excited) ${\cal G}$.  We believe that this reaction is
inherently interesting because there is little conventional background,
so that with any appreciable rate the residual material accompanying a
final $\Upsilon$ is likely to be remarkable in significant ways.
Furthermore, this sort of search would be very much in the spirit of
crossing symmetry:  Well separated heavy quarks manifestly feel the
confinement force.  Therefore these quarks are the best candidates to
radiate the associated meson, but in processes which involve large
distance scales, such as descent from one quarkonium state to another,
rather than annihilation, which is intrinsically short-distance in
character.

This work was supported in part by the Department of Energy and in part
by the National Science Foundation.  We thank Howard Georgi, David
Kreinick, Edward Shuryak and Jacobus Verbaarschot for comments.

\pagebreak

\end{document}